\documentclass{aip-cp}

\usepackage[numbers]{natbib}
\usepackage{graphicx}
\usepackage{bm}
\usepackage{xcolor}
\usepackage{subcaption}
\usepackage{url}

\newcommand{\braket}[1]{{\langle}{#1}{\rangle}}

\newcommand{\ket}[1]{{|}{#1}{\rangle}}

\begin{document}

\title{Quasiparticle-vibration coupling effects on nuclear transitions of astrophysical interest}

\author[aff1,aff2,aff3]{Caroline Robin\corref{cor1}}
\eaddress[url]{http://www.aip.org}
\author[aff1,aff4]{Elena Litvinova}
\eaddress{elena.litvinova@wmich.edu}

\affil[aff1]{Department of Physics, Western Michigan University, Kalamazoo, MI 49008, USA.}
\affil[aff2]{Institute for Nuclear Theory, University of Washington, Seattle, WA 98195, USA.}
\affil[aff3]{JINA-CEE, Michigan State University, East Lansing, MI 48824, USA.}
\affil[aff4]{National Superconducting Cyclotron Laboratory, Michigan State University, East Lansing, MI 48824, USA.}
\corresp[cor1]{Corresponding author: carolr8@uw.edu}

\maketitle

\begin{abstract}
The relativistic quasiparticle time-blocking approximation (RQTBA) is applied to the description of nuclear excitation modes of astrophysical interest. This method is based on the meson-nucleon Lagrangian and goes beyond the standard relativistic quasiparticle random-phase approximation (RQRPA) by treating the coupling between single quasiparticles and collective vibrations of the nucleus. We calculate electric dipole transitions and Gamow-Teller modes in the (p,n) direction in a few Sn isotopes and obtain the rates of (n,$\gamma$) reaction and $\beta^-$-decay processes, which govern the r-process nucleosynthesis, in a unified RQTBA framework. 
Gamow-Teller transitions in the (n,p) branch, which in principle can serve for the modeling of stellar evolution, are also investigated, and $^{90}$Zr is taken as a study case.
\end{abstract}

\section{INTRODUCTION}
The atomic nucleus is a complex system at the frontiers between microscopic and macroscopic worlds. As it is made of interacting nucleons, it can display properties common to various quantum many-body systems. For instance, nuclei present evidences of an underlying shell structure, they can decay via particle emission or radiation, and exhibit collective phenomena, such as superfluidity, vibrations or rotations. Additionally, nuclei have specific features. For example, they are made of two types of particles that are protons and neutrons, which makes them sensitive to at least three of the four fundamental interactions and leads to the appearance of interesting phenomena such as \textit{e.g.} pairing between protons and neutrons, the formation of a neutron skin in neutron-rich nuclei, as well as the possibility for charge-exchange or weak-interaction processes.
\\
Despite tremendous progress in the theoretical description of nuclei, and a large number of theoretical methods developed over the years, the consistent treatment of complex correlations between nucleons remains a challenge.
While growing amount of experimental data allows for testing and benchmarking theoretical many-body methods, a large fraction of existing nuclei remain too unstable to be experimentally reached in the near future. 
The modeling of astrophysical processes requires the knowledge of excitation spectra, decay and reaction rates for a large number of nuclei, especially those at the drip-lines and, therefore, relies on theoretical models which should be as universal, consistent, precise and predictive as possible.
As an example, the r-process nucleosynthesis, which consists of an alternation of beta-decay and neutron capture, runs through extremely neutron-rich nuclei, and requires a consistent description of electromagnetic and spin-isospin transitions, first of all, electric dipole and charge-exchange Gamow-Teller in the (p,n) channel, in mid-mass to heavy nuclei. Gamow-Teller transitions in the (n,p) branch also provide electron-capture rates that are a necessary ingredient for simulating the late evolution of massive stars, during which nuclei with large neutron excess can appear.
In this context, methods based on the density functional theory (DFT), or mean field approximation, such as the (quasiparticle) random phase approximation ((Q)RPA) \cite{RingSchuck} are largely used as they are applicable to a wide range of masses. Such methods, however, typically suffer from a lack of inter-nucleon correlations in their description of nuclei. 
\\
In this work we apply the relativistic quasiparticle time-blocking approximation (RQTBA), which has been developed during the last decade (see \textit{e.g.} References \cite{Litvinova2007,Litvinova2008,Marketin2012,Robin2016}), to the description of various excitation modes of astrophysical interest. 
Based on the meson-nucleon Lagrangian of quantum hadrodynamics, the present approach starts from a covariant DFT description of the nucleus and uses Gorkov-Green function techniques and nuclear field theory to introduce correlations and compute the response of nuclei to various external fields.
While the (relativistic) QRPA ((R)QRPA) describes nuclear vibrations as superpositions of one-particle-one-hole (1p-1h), or two-quasiparticle (2 qp), excitations on top of the ground state, RQTBA goes a step further by considering the coupling between single nucleons and collective vibrations of the nucleus, which introduces complex 2p-2h, or 4 qp, configurations in the 2qp$\otimes$phonon coupling scheme.
\\
This paper is organized as follows. In the first section we briefly remind the formalism of the RQTBA in both like-particle and charge-exchange channels. In the second section we apply the method to the calculation of electric dipole modes and Gamow-Teller transitions in the (p,n) channels, in a few Sn isotopes situated on the r-process path. The resulting strength functions then serve for the calculation of neutron-capture rates and beta-decay half-lives. We also investigate the (n,p) branch of the Gamow-Teller modes in $^{90}$Zr as a test case for this type of nuclear response having in mind future applications to calculations of electron capture rates.
Finally, we give a summary and conclusion to this work.

\section{FORMALISM}
\label{sec:formalism}
\begin{figure}[h]
\centerline{\includegraphics[width=400pt]{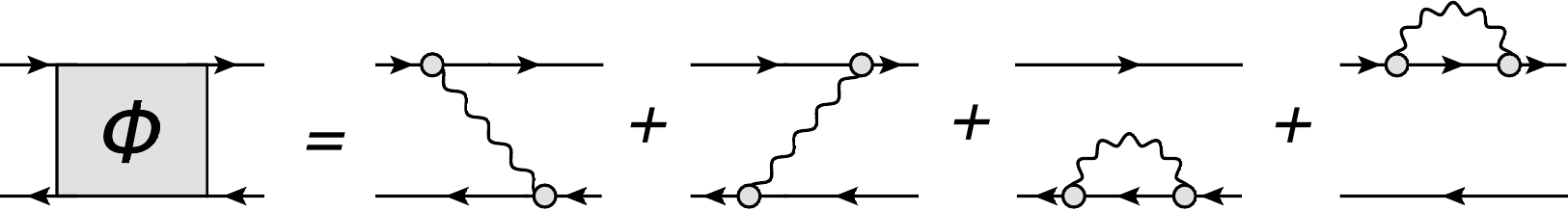}}
\caption{Feynman diagram representation of the dynamical quasiparticle-vibration interaction $\Phi(\omega)$. The straight lines represent nucleon quasiparticle states while the wiggly lines represent a nuclear vibration (phonon). The gray circles denote quasiparticle-vibration coupling vertices.}
\label{fig:QVC}
\end{figure}

When studying the response of a nucleus to an external field characterized by a one-body transition operator $\hat O$, the corresponding strength distribution can be directly computed from the knowledge of the so-called response function $R$, or propagator of two correlated nucleons in the medium, as
\begin{eqnarray}
S(E) &=& \sum_f  |\braket{\Psi_f | \hat O| \Psi_i}|^2 \delta(E-E_f+E_i) -  |\braket{\Psi_f | \hat O^\dagger | \Psi_i}|^2 \delta(E+E_f-E_i)\; , \nonumber \\
         &=& \frac{-1}{\pi} \lim_{\Delta \rightarrow 0^+} \mbox{Im} \braket{\Psi_i | \hat O^\dagger \hat R(E+i\Delta) \hat O  | \Psi_i} \; ,
\label{eq:strength}         
\end{eqnarray}
where $\ket{\Psi_i}$ and $\ket{\Psi_f}$ denote the initial and final states of the transition, with energies $E_i$ and $E_f$, respectively.
The nature of the external field under study (\textit{e.g.} like-particle, charge-exchange, particle-number conserving or two-particle transfer), selects a particular channel of the response function and provides spectra and transition probabilities either in the same $(N,Z)$ nucleus or in neighboring $(N \pm 1, Z \mp 1)$, $(N \pm 2, Z)$,  $(N, Z \pm 2)$ nuclei.\\
In this work we will focus on A-conserving like-particle and charge-exchange channels, which are relevant for electric dipole and Gamow-Teller transitions.
\\ \\
According to the formalism of the relativistic quasiparticle time-blocking approximation \cite{Litvinova2008,Robin2016} we solve the Bethe-Salpeter equation (BSE) for the response function in the corresponding channels.
\\
In the like-particle case, the final BSE reads \cite{Litvinova2008}
\begin{eqnarray}
R^{\eta_1 \eta_4 , \eta_2 \eta_3}_{k_{1_\tau} k_{4_{\tau}}, k_{2_\tau} k_{3_{\tau}}} (\omega) = \widetilde{R}^{(0) \eta_1 \eta_4 , \eta_2 \eta_3}_{k_{1_\tau} k_{4_{\tau}}, k_{2_\tau} k_{3_{\tau}}} (\omega)
+ \sum_{k_{5_{\tau}} k_{6_{\tau}} k_{7_{\tau}} k_{8_{\tau}}} \sum_{ \eta_5 \eta_6 \eta_7 \eta_8} \widetilde{R}^{(0) \eta_1 \eta_6 , \eta_2 \eta_5}_{k_{1_\tau} k_{6_{\tau}}, k_{2_\tau} k_{5_{\tau}}} (\omega) 
W^{\eta_5 \eta_8 , \eta_6 \eta_7}_{k_{5_{\tau}} k_{8_{\tau}}, k_{6_{\tau}} k_{7_{\tau}}} (\omega) R^{\eta_7 \eta_4 , \eta_8 \eta_3}_{k_{7_{\tau}} k_{4_{\tau}}, k_{8_{\tau}} k_{3_{\tau}}} (\omega) \; , \nonumber \\ 
\label{eq:BSE_lp}
\end{eqnarray}
where $\omega = E+i\Delta$ is the energy variable, $k_{i_\tau}$ denotes a set of nucleonic quantum numbers with isospin projection $\tau$, and in open-shell nuclei with superfluid pairing correlations $\eta_i=\pm$ denotes the upper and lower component in the Nambu-Gorkov space \cite{Nambu,Gorkov}. 
In Equation (\ref{eq:BSE_lp}), $\widetilde{R}^{(0)} (\omega)$ is the propagator of two nucleons evolving freely in the mean field, and $W$ is the effective nucleon-nucleon interaction 
\begin{equation}
W (\omega) = \widetilde{V}_{\sigma} + \widetilde{V}_{\omega} + \widetilde{V}_{\rho} + \widetilde{V}_{\gamma}  + \Phi(\omega) - \Phi(0) \; ,
\label{eq:neutral_int}
\end{equation}
given by the sum of the static meson ($\sigma$, $\omega$, $\rho$) and photon ($\gamma$) exchanges, and a quasiparticle-vibration coupling (QVC) energy-dependent term $\Phi(\omega)$ which is responsible for damping effects.
The latter describes the virtual emission and re-absorption of a nuclear vibration ("phonon") by a quasiparticle, as well as the exchange of a phonon between two quasiparticles (see Figure \ref{fig:QVC}). In Equation (\ref{eq:neutral_int}), the QVC interaction at zero excitation energy $\Phi(0)$ is subtracted in order to avoid double counting of effects implicitly incorporated in the parameters of the phenomenological meson-exchange interaction, see References \cite{Lalazissis1997,Lalazissis2009}. A precise expression of $W (\omega)$ is obtained in the framework of the relativistic nuclear field theory and can be found in Ref. \cite{Litvinova2008}.
\\ \\
In the charge-exchange channel the BSE for the proton-neutron response function reads \cite{Robin2016}
\begin{eqnarray}
R^{\eta_1 \eta_4 , \eta_2 \eta_3}_{k_{1_p} k_{4_n}, k_{2_n} k_{3_p}} (\omega) = \widetilde{R}^{(0) \eta_1 \eta_4 , \eta_2 \eta_3}_{k_{1_p} k_{4_n}, k_{2_n} k_{3_p}} (\omega)
+ \sum_{k_{5_p} k_{6_n} k_{7_p} k_{8_n}} \sum_{ \eta_5 \eta_6 \eta_7 \eta_8} \widetilde{R}^{(0) \eta_1 \eta_6 , \eta_2 \eta_5}_{k_{1_p} k_{6_n}, k_{2_n} k_{5_p}} (\omega) 
W^{(pn)\eta_5 \eta_8 , \eta_6 \eta_7}_{k_{5_p} k_{8_n}, k_{6_n} k_{7_p}} (\omega) R^{\eta_7 \eta_4 , \eta_8 \eta_3}_{k_{7p} k_{4_n}, k_{8_n} k_{3_p}} (\omega) \; , \nonumber \\ \label{eq:BSE_pn}
\end{eqnarray}
where $k_{i_p}$ and $k_{i_n}$ denote proton and neutron single-particle states, respectively.
In this channel only the isovector mesons (limited to pion and rho-meson in this work) contribute to the static part of the interaction:
\begin{equation}
W^{(pn)} (\omega) = \widetilde{V}_{\rho} + \widetilde{V}_{\pi} + \widetilde{V}_{\delta_\pi} + \Phi^{(pn)}(\omega) \, (- \Phi^{(pn)}(0) )\; .
\label{eq:pn_int}
\end{equation}
In Equation (\ref{eq:pn_int}), $V_{\delta_\pi}$ is the zero-range Landau-Migdal term \cite{Bouyssy1987}, taken in this work with strength parameter $g'=0.6$, as the Fock term is not treated explicitly \cite{Liang2008}.
The proton-neutron QVC interaction $\Phi^{(pn)}(\omega)$ takes a similar form as in the like-particle channel. An analytical expression for $\Phi^{(pn)}(\omega)$ can be found in Ref. \cite{Robin2016}.
In the case of unnatural parity modes, such as Gamow-Teller transitions, the $\rho$-meson has a very minor contribution so that the pion is the only meson that contributes non-negligibly to the transition.
Since the pion does not contribute at the Hartree level, due to parity conservation, it is not adjusted at the mean field level, and is considered here with the free coupling constant $\frac{f_\pi^2}{4\pi}=0.08$. Therefore, no subtraction is made in the case of unnatural parity modes. However, for natural parity modes, such as Fermi transitions, only the rho meson contributes and a subtraction is made in this case.

\section{APPLICATIONS}
\subsection{Transitions relevant to the r-process nucleosynthesis: neutron-capture and beta-decay rates in neutron-rich Sn isotopes} \label{sec:rproc}
\begin{figure}[h!]
\centerline{\includegraphics[width=450pt]{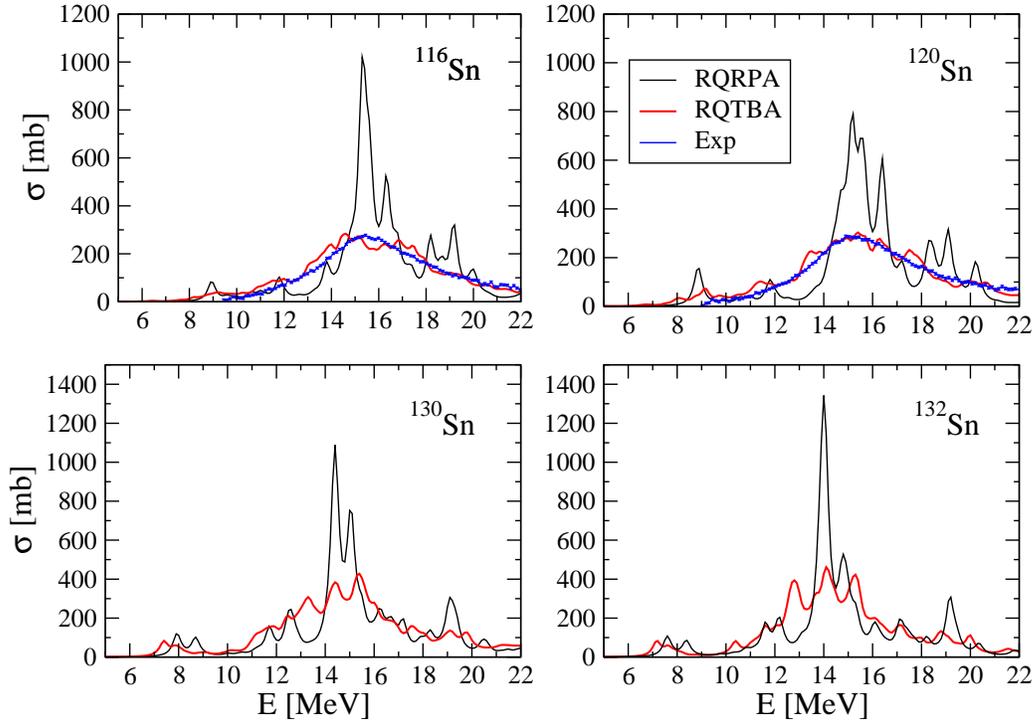}}
\caption{E1 photo-absorption cross-sections in $^{116}$Sn, $^{120}$Sn, $^{130}$Sn and $^{132}$Sn calculated in RQRPA (black) and RQTBA (red) and compared to the experimental data (blue) \cite{Lepetre}.}
\label{fig:E1-cross}
\end{figure}

\begin{figure}[h!]
\centerline{\includegraphics[width=350pt]{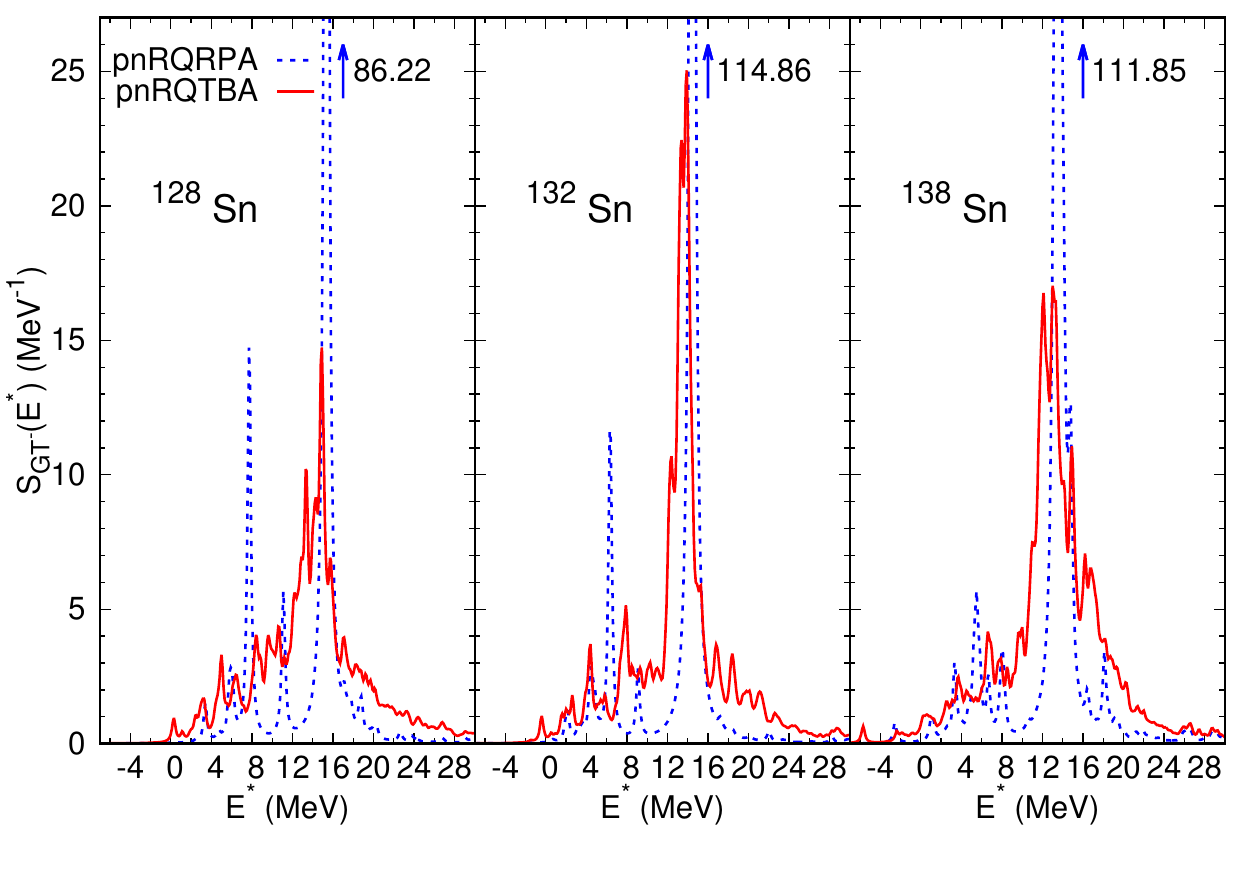}}
\caption{GT$^-$ strength in $^{128}$Sn, $^{132}$Sn and $^{138}$Sn calculated in pn-RQRPA (dashed blue curves) and in pn-RQTBA (plain red). The distributions have been obtained using a smearing $\Delta=200$ keV. The excitation energy $E^*$ is taken with respect to the ground state of the parent Sn nuclei.}
\label{fig:str_Sn}
\end{figure}

\begin{figure}[h!]
\centerline{\includegraphics[width=250pt]{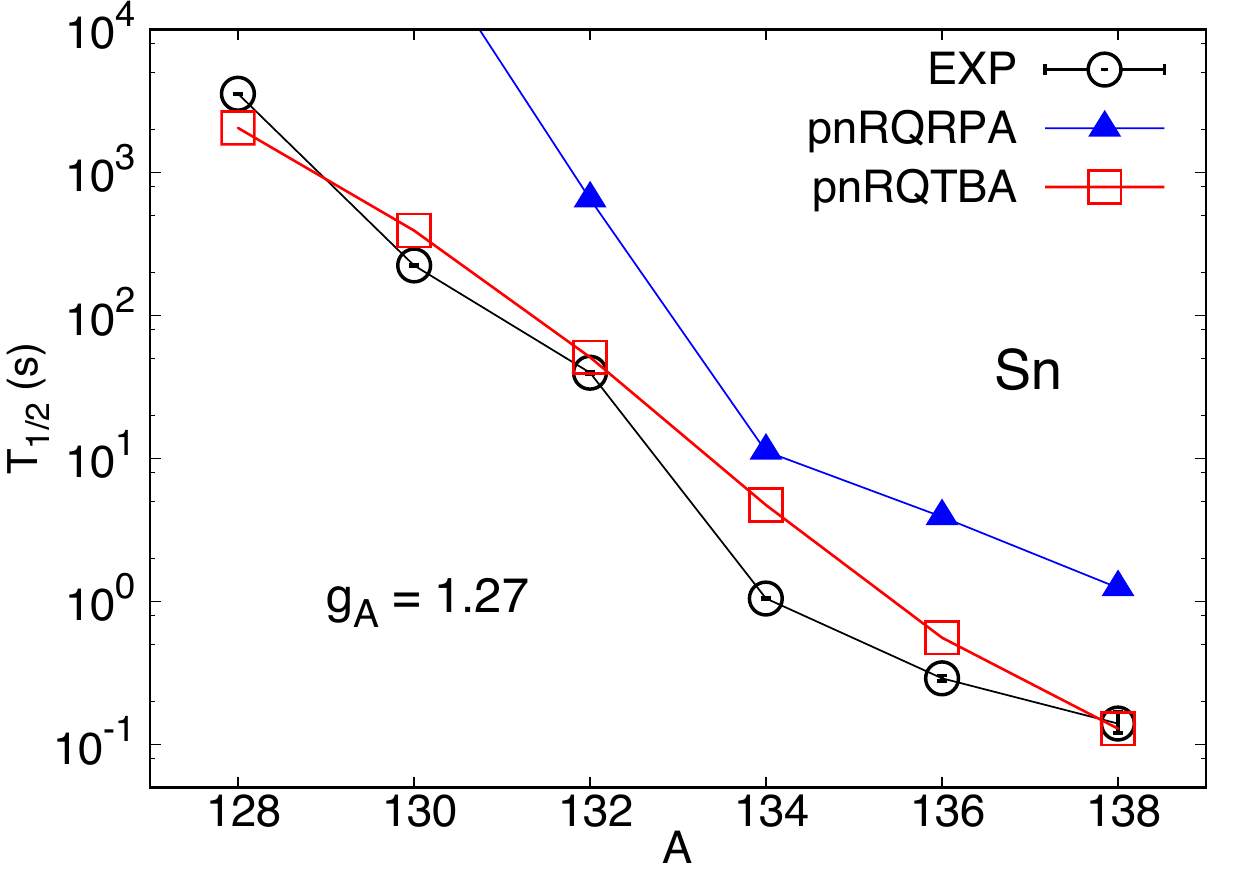}}
\caption{Half-lives of $128 \leq A \leq 138$ even-even Sn isotopes, in seconds. They have been obtained using the bare value of the weak-axial coupling constant $g_A \simeq 1.27$.
The experimental values are taken from \cite{nndc}.}
\label{fig:h_lives}
\end{figure}

\begin{figure}[h!]
\centerline{\includegraphics[width=350pt]{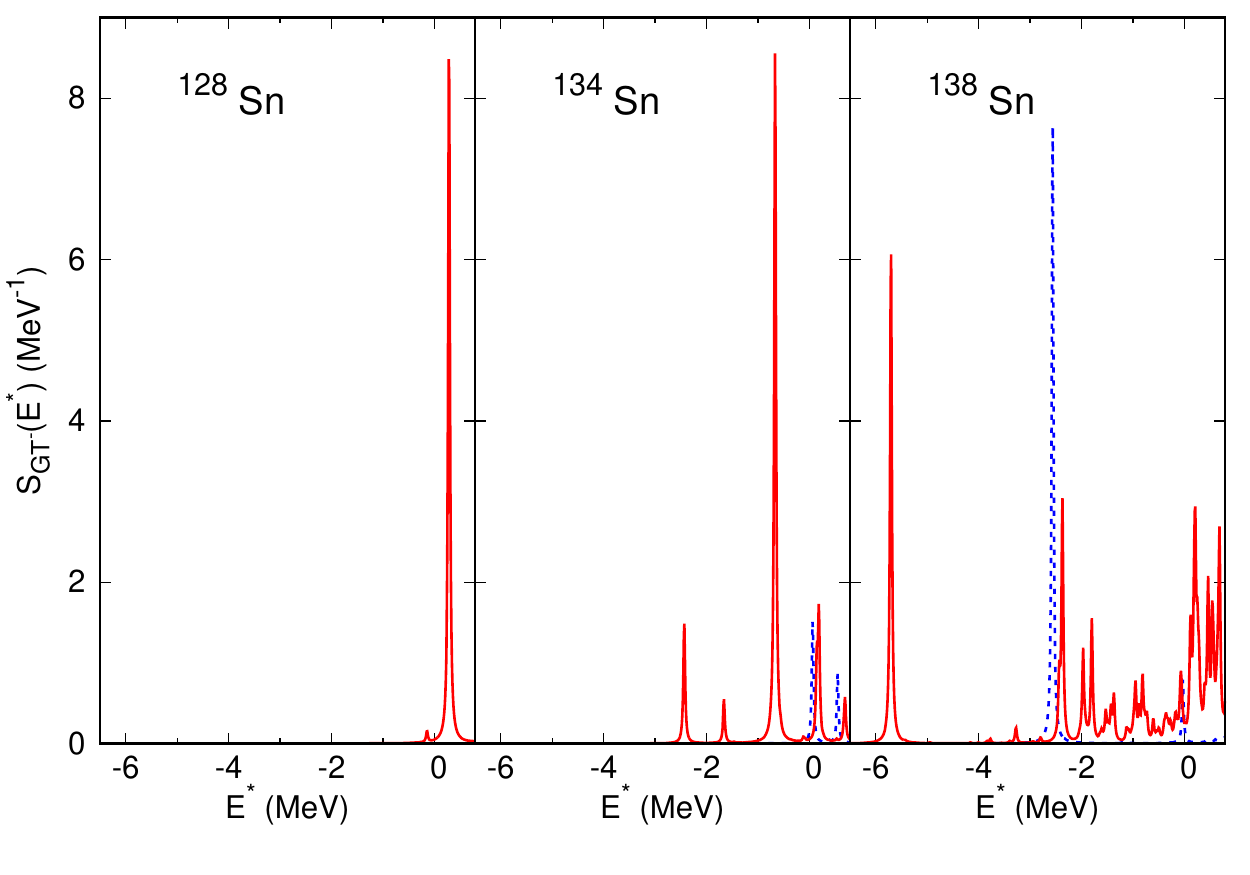}}
\caption{Low-energy GT$^-$ strength contributing to the $\beta^-$-decay half-lives in $^{128}$Sn, $^{134}$Sn and $^{138}$Sn calculated in pn-RQRPA (dashed blue curves) and in pn-RQTBA (plain red). The distributions have been obtained using a smearing $\Delta=20$ keV. The excitation energy $E^*$ is taken with respect to the ground state of the parent Sn nucleus.}
\label{fig:str_Sn_Qb}
\end{figure}

Astrophysical simulations of the r-process require a consistent description of like-particle electric dipole (E1) transitions and charge-exchange Gamow-Teller (GT) modes, which give the leading contributions to the neutron capture and beta decay rates. The Sn region around $N=82$ is particularly important for r-process studies as $^{132}$Sn is a so-called "waiting-point" nucleus where neutron-capture and photodesintegration processes are in equilibrium and $\beta^-$-decay favorably occurs.
Here we calculated E1 modes and GT transitions in the (p,n) branch (GT$^-$ transitions) in a few Sn isotopes using the following numerical scheme: as a first step, the basis $(k_{i_\tau},\eta_i)$ of single quasiparticles is obtained by performing a relativistic mean-field calculation using NL3 parametrization of the meson-exchange interaction \cite{NL3, Lalazissis1997} and monopole-monopole pairing force, as described in Reference \cite{Litvinova2008}, for which the Hartree-Bogoliubov approximation reduces to the Hartree+Bardeen-Cooper-Schrieffer (Hartree+BCS) one. The spectrum of collective phonons that are coupled to the quasiparticles are calculated within the RQRPA based on the same NL3 parameter set. In these applications we select isoscalar phonons with angular momentum and parity $2^+,3^-,4^+,5^-,6^+$ which realize at least $5\%$ of the higher transition probability for a given multipolarity, and with an excitation energy of up to $15$ MeV. The contribution of other types of vibrations (isovector phonons, phonons with unnatural parities, non-collective or higher-energy vibrations) are expected to be small as investigated in \textit{e.g.} Ref. \cite{subtraction,Robin2016,RobinINPC}. Finally, we solve Equations (\ref{eq:BSE_lp}) and (\ref{eq:BSE_pn}) for a given multipolarity $J^\pi$. 
In order to appreciate the effect of the coupling between single-particle and collective degrees of freedom, in the following applications we compare the results obtained at the RQRPA level (\textit{i.e.} neglecting the dynamical QVC interaction $\Phi(\omega)$ in Equations (\ref{eq:BSE_lp}) or $\Phi^{(pn)}(\omega)$ in Equation (\ref{eq:BSE_pn})) and at the RQTBA level (\textit{i.e.} solving the full BSE (\ref{eq:BSE_lp}) or (\ref{eq:BSE_pn})). In the latter case, due to numerical limitations, the energy window, in which the QVC effects are included, is limited to $25$ MeV around the Fermi level.
\\
\\
\\
\indent We show in Figure \ref{fig:E1-cross} the dipole photo-absorption cross sections in $^{116}$Sn, $^{120}$Sn, $^{130}$Sn and $^{132}$Sn calculated in both RQTBA (red lines) and RQRPA approximation (black lines) as \cite{Litvinova2009}
\begin{equation}
\sigma_{E_1}(E_\gamma) = \frac{16 \pi^3 e^2}{9 \hbar c} E_\gamma S_{E_1}(E_\gamma) \; ,
\end{equation}
where $E_\gamma$ denotes the photon energy and $S_{E_1}(E_\gamma)$ is the E1 strength distribution which was obtained here using a smearing parameter $\Delta=200$ keV in Equation (\ref{eq:strength}). 
The QVC effects introduced in the RQTBA lead to a nice agreement with the available experimental data (blue) in $^{116}$Sn, $^{120}$Sn \cite{Lepetre}. Comparison to available data for $^{130,132}$Sn is also very reasonable, but can be only made indirectly, see Reference \cite{Litvinova2009-2}.
$\;$\\ \\
\begin{figure}[ht!]
\centerline{\includegraphics[width=400pt]{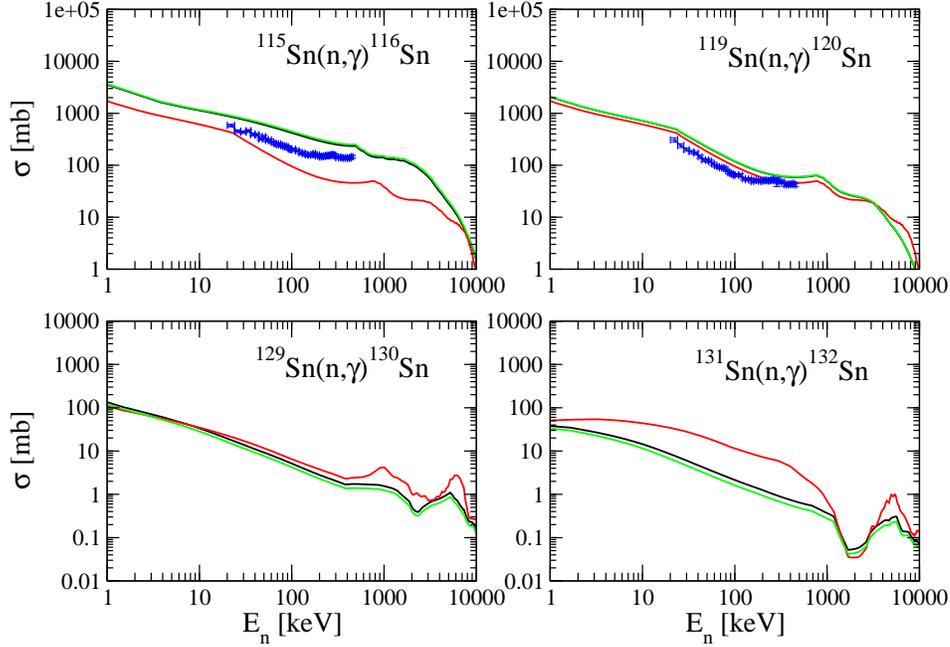}}
\caption{Cross-sections for $(n,\gamma)$ reactions on $^{116}$Sn, $^{120}$Sn, $^{130}$Sn and $^{132}$Sn, calculated within the present RQTBA (red lines), and with Lorentzian parametrizations of References \cite{Thielemann} (green) and \cite{Lorentzian1} (black) of the dipole strength. The experimental data \cite{Lepetre} is shown with blue error bars. }
\label{fig:xs}
\end{figure}
In order to obtain neutron-capture reaction rates, the $(n,\gamma)$ cross sections were then calculated using the dipole photo-absorption cross sections of Figure \ref{fig:E1-cross} within the Hauser-Feshbach statistical model \cite{Hauser} using the Brink-Axel hypothesis \cite{Brink,Axel}, as described in details in Reference \cite{Litvinova2009}.
In Figure \ref{fig:xs} we show the resulting $(n,\gamma)$ cross sections for $^{116}$Sn, $^{120}$Sn, $^{130}$Sn and $^{132}$Sn compared to data.
\begin{figure}[h!]
\centerline{\includegraphics[width=400pt]{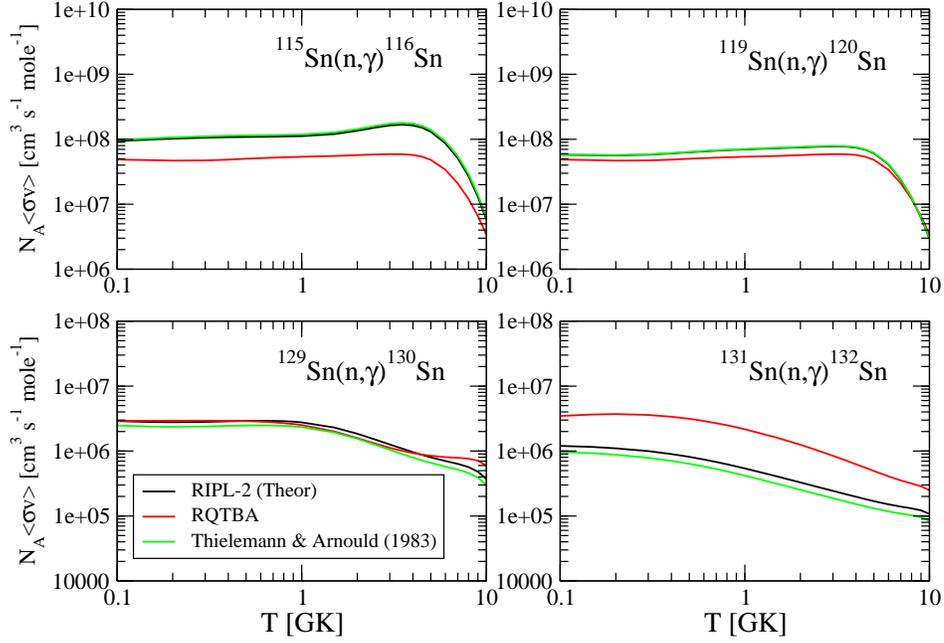}}
\caption{(n,$\gamma$) reaction rates calculated from the cross sections of Figure \ref{fig:xs} as a function of temperature \cite{Litvinova2009}.}
\label{fig:ng_rates}
\end{figure}
As standard r-process simulations often use a gamma strength function $S_{E_1}(E_\gamma)$ approximated by a Lorentzian distribution with parameters adjusted to experimental data or theoretical systematics, we compare our microscopic RQTBA results (red lines) to the ones obtained with two Lorentzian parametrizations of the strength. 
The black curves (RIPL-2) use Lorentzian centroid, peak cross section and width extracted from data and given by Reference \cite{Lorentzian1}, while the green curves (Thielemann and Arnould, 1983) use the prescription given in Reference \cite{Thielemann}. 
One sees that for the isotopes with smaller neutron number the RQTBA cross sections lie below the ones obtained with the Lorentzian models which typically predict a larger low-energy strength. As the neutron number increases, the pygmy resonance, which is an oscillation of the neutron skin against the core of the nucleus, clearly appears in the microscopic strength distribution below 10 MeV (see Figure \ref{fig:E1-cross}) and moves closer to the neutron threshold. This enhancement of the low-energy strength then leads to a much larger cross section predicted by RQTBA in comparison to the phenomenological models. The importance of the pygmy resonance on the description of neutron-capture cross sections was first pointed out in Reference \cite{Goriely}.
\\  
\\
In Figure \ref{fig:ng_rates} we show the neutron capture rates as a function of the temperature in a range relevant for the r-process. These rates have been calculated by averaging the above cross sections multiplied by the neutron velocity with Maxwell distribution (see \textit{e.g.} Reference \cite{ng-rates}).
The conclusion drawn for the cross sections applies to the neutron-capture rates. In $^{132}$Sn the rates calculated within RQTBA are larger by a factor $\sim 5$ as compared with the standard Lorentzian models. This clearly demonstrates the importance of microscopic effects on the calculation of the rates, which should consequently impact r-process simulations.
\\ \\ \\
\indent In Figure \ref{fig:str_Sn} we show the (p,n) branch of the Gamow-Teller strength distributions (GT$^-$) in $^{128}$Sn, $^{132}$Sn and $^{138}$Sn. We show with dashed blue lines the results obtained at the proton-neutron RQRPA level (pn-RQRPA) (neglecting the dynamical interaction $\Phi^{(pn)}(\omega)$ in Equation (\ref{eq:pn_int})), and with plain red lines the results obtained at the pn-RQTBA level, i.e. when taking into account QVC effects. These strength distributions have been obtained using the (unquenched) Gamow-Teller operator $\hat{O}_{GT^-} = \sum_i \bm{\Sigma}(i) \tau_-(i)$, and using an imaginary parameter $\Delta=200$ keV for the energy argument in Equation (\ref{eq:strength}).
Similarly to the like-particle case, the main consequence of coupling single-nucleons to collective phonons is a strong fragmentation and spreading of the transition strength over a larger energy range, including both the low- and high-energy regions.
$\;$ \\
In order to investigate the effect of QVC on $\beta^-$-decay rates we have calculated $\beta^-$-decay half-lives in the allowed Gamow-Teller approximation as
\begin{equation}
T_{1/2}^{-1} = \frac{g_A^2}{D} \int_{\Delta B}^{\Delta_{nH}} f(Z,\Delta_{np}-E) S_{GT^-}(E) dE \; ,
\label{eq:h_lives}
\end{equation}
where $D=6163.4$ s, and $\Delta_{nH}=0.78227$ MeV and $\Delta_{np}=1.293$ MeV are the mass differences between the neutron and the Hydrogen atom and the neutron and the proton respectively. $\Delta B = B(A,Z)- B(A,Z+1)$ is the binding energy difference between the parent and daughter nuclei, 
and $f$ is the leptonic phase-space factor \cite{Bahcall1966}. Finally, we use the bare value of the weak axial coupling constant $g_A \simeq 1.27$ \cite{gA}. The results are shown in Figure \ref{fig:h_lives}.
In order to obtain the half-lives to a good accuracy, we have used in Equation (\ref{eq:h_lives}) the GT$^-$ strength smeared with a small imaginary parameter: $\Delta=20$ keV.
As examples, we show in Figure \ref{fig:str_Sn_Qb} the corresponding low-energy strengths in $^{128}$Sn, $^{134}$Sn and $^{138}$Sn.
Overall, pn-RQRPA systematically overestimates the half-lives of Sn isotopes by at least an order of magnitude, while the QVC included in pn-RQTBA improves the results essentially.
In $^{128}$Sn and $^{130}$Sn, the pn-RQRPA does not predict any state within the integration limits and, therefore, describes these isotopes as stable, in contradiction with experimental data. The downward shift of the low-lying states due to QVC effects in the pn-RQTBA leads to the appearance of states in the integration window and to a finite lifetime for these two nuclei. As the neutron excess increases, both approximations predict an increase of the low-energy strength and a decrease of the half-lives. However, the pn-RQTBA systematically predicts a larger amount of states at lower energies, for which the leptonic phase-space in Equation (\ref{eq:h_lives}) takes greater values. Similar conclusions were obtained for Ni isotopes in Reference \cite{Robin2016}.

\subsection{Nuclear excitations relevant to stellar evolution: GT$^+$ transitions modes} \label{sec:GT+}
\begin{figure}[h]
\centerline{\includegraphics[width=300pt]{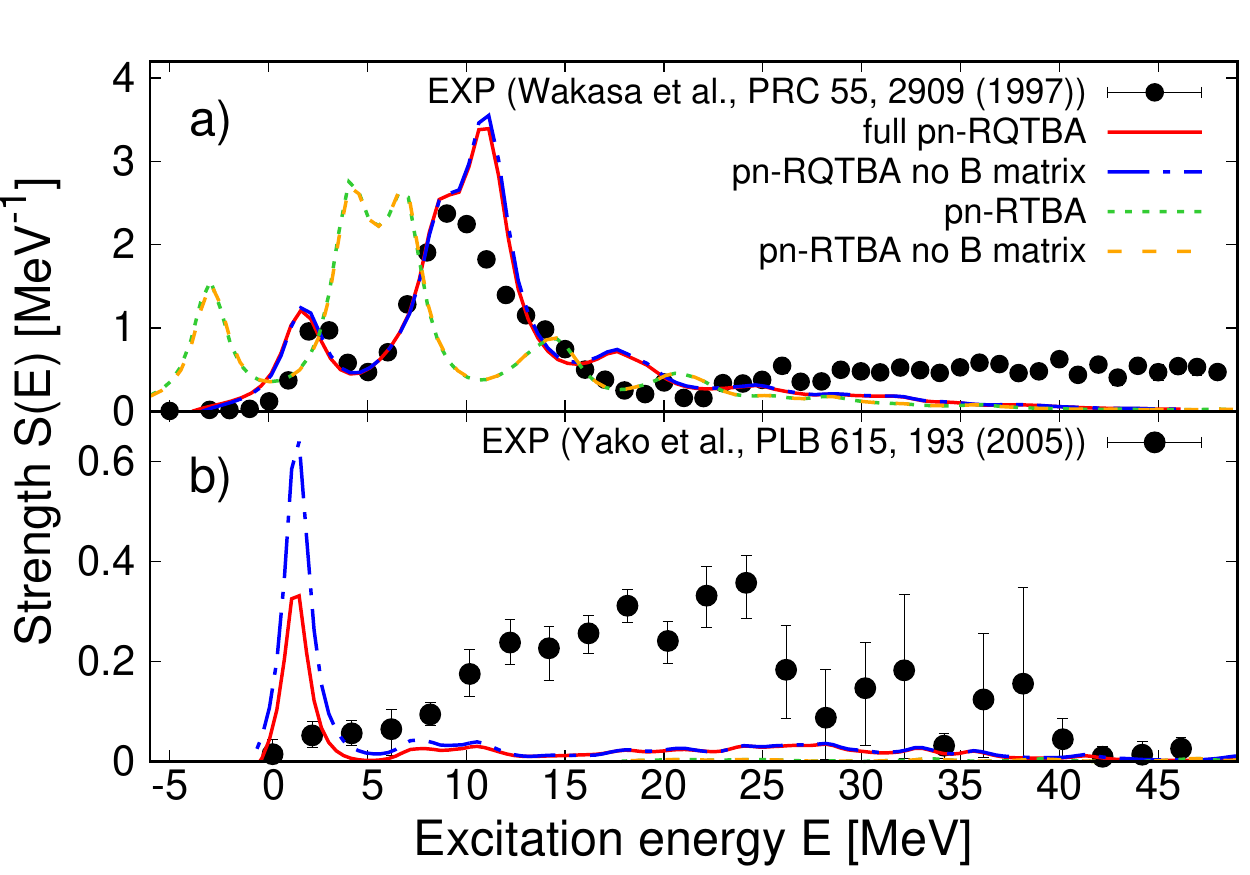}}
\caption{a): GT$^-$ strength distribution in $^{90}$Zr within full pn-RQTBA (plain red), pn-RQTBA without B-matrix (dashed blue), pn-RQTBA without pairing in the mean field (pn-RTBA) (dashed green), pn-RQTBA without pairing in the mean field and without B matrix (dashed orange). The strength has been smeared with $\Delta=1$ MeV to match the experimental resolution. The data has been extracted from \cite{Wakasa1997}. b): same as a) for the GT$^+$ distribution in $^{90}$Zr. Here the strength has been smeared with $\Delta=750$ MeV to match the resolution of the experimental data extracted from \cite{Yako2005}.}
\label{fig:str_90Zr}
\end{figure}

While GT transitions in the (p,n) channel set the time scale for the r-process nucleosynthesis, GT modes in the (n,p) branch (GT$^+$) are in direct relation with electron-capture rates that are a major ingredient in the evolution of stars that become supernovae. In this context, electron capture on $pf$- or $sdg$-shell nuclei are crucial.
In a pure independent-particle picture, in nuclei with $N>Z$, transitions in the (n,p) channel are suppressed due to Pauli blocking.
Thus, the presence of correlations in the ground state strongly influences the GT$^+$ strength distribution in these nuclei.
Apart from superfluid pairing correlations, which can be included with a sufficient accuracy at the mean field level (here within the Hartree+BCS theory), beyond-mean-field ground-state correlations (GSC), in the language of Green functions techniques, are typically induced by "backward-going" Feynman diagrams. In the (R)QRPA, these correspond to the well known "B-matrix" \cite{RingSchuck} which is responsible for allowing transitions from particle states (empty states in a pure independent-particle picture) to hole states (occupied states in a pure independent-particle picture). At the present level of approximation of the RQTBA method, so-called "resonant approximation", we only introduce forward-going diagrams induced by the coupling to phonons (see Figure \ref{fig:QVC}). Therefore, the only GSC that are included in this work are the ones due to pairing and RQRPA B-matrix. 
\\
\\
Although not of direct astrophysical relevance, the nucleus $^{90}$Zr is a good study case as it has been measured in both (p,n) and (n,p) channels \cite{Yako2005,Wakasa1997}, and is known to have weak pairing effects due to the sub-shell closure at N=40.
In order to evaluate the importance of GSC, we have calculated GT$^-$ and GT$^+$ strength distributions in this nucleus at different levels of approximations: in the pn-RQTBA including and excluding pairing correlations, and including and excluding the RQRPA B-matrix. Again the calculations have been done using the unquenched GT$^+$ and GT$^-$ transition operators, and according to the numerical scheme exposed in the previous section. Here, however, the smaller number of nucleons allowed us to couple quasiparticles to phonons with an excitation energy up to $20$ MeV, and to extend the QVC window up to $30$ MeV around the Fermi level. 
The results are shown in Figure \ref{fig:str_90Zr} and compared to the experimental data. To that aim, the strengths have been folded with a smearing parameter corresponding to the experimental resolution \textit{i.e.} $\Delta=750$ keV and $\Delta=1$ MeV for the GT$^+$ and GT$^-$ channels, respectively.
$\;$ \\
In the GT$^-$ channel, as expected, superfluid pairing correlations induce an upward shift of the strength but do not importantly modify the shape of the spectrum. 
The inclusion of the B-matrix has almost no effect on the transitions. Globally, the pn-RQTBA agrees quite well with the data with or without residual (beyond-mean-field) GSC.
The conclusions are, however, different in the GT$^+$ channel. First of all, the presence of superfluid pairing correlations is solely responsible for the appearance of strength in this branch, as the latter drops to almost zero when such correlations are turned off (the strength is barely visible on the figure in that case). Pairing correlations smear the occupations of the single-particle states around the Fermi level and unblock the transition from the $\pi g_{9/2}$ to $\nu g_{7/2}$ which leads to the appearance of the GT$^+$ state around 2 MeV.
When pairing is turned on, the B-matrix leads to a quite noticeable quenching of this transition. 
Comparing to the data, we note that the pn-RQTBA strength still does not agree with the experimental trend. In order to investigate this issue further, we are currently working on the inclusion of additional GSC, in particular, those induced by QVC \cite{Robin2017}. We expect that such effects could potentially improve the description of the GT$^+$ strength, and consequently the description of electron-capture rates, in neutron-rich nuclei.

\section{SUMMARY AND CONCLUSION} \label{sec:conclu}
In this work we have applied the RQTBA approach to the description of various excitation modes in nuclei. As it is now available in both the like-particle and proton-neutron channels, this approach is able to provide a consistent and microscopic description of both alternating phases of the r-process nucleosynthesis that are $\beta^-$-decay and neutron-capture. In this study we showed that microscopic effects are important for the calculation of $(n,\gamma)$ rates, especially in very unstable neutron-rich nuclei, in which the pygmy dipole resonance is close to the neutron threshold. Additionally, QVC effects, which are missing in (R)QRPA, are found to be crucial for a precise description of the low-energy GT$^-$ strength and for adequate predictions of the $\beta^-$-decay half-lives. The calculation of the GT$^+$ transition strength in $^{90}$Zr revealed a disagreement with the experimental data, while the GT$^-$ distribution was quite nicely reproduced. Such a disagreement could potentially be due to the lack of complex ground-state correlations in the present implementation of the RQTBA, as such correlations are crucial for GT$^+$ (GT$^-$) transitions in $N>Z$ ($N<Z$) nuclei. In order to improve the description of such transitions we are currently working on the inclusion of GSC induced by the QVC effects \cite{Robin2017}.

\section{ACKNOWLEDGMENTS}
The authors thank Hans Peter Loens for sharing his results on the radiative neutron capture cross sections and reaction rates calculation. This work was supported by US-NSF Grant PHY-1404343 and US-NSF Career Grant 1654379. 
Support by the Institute for Nuclear Theory under US-DOE Grant DE-FG02-00ER41132, and by JINA-CEE under US-NSF Grant PHY-1430152 are also acknowledged.

\nocite{*}
\bibliographystyle{aipnum-cp}%
\bibliography{NMP17}%

\end{document}